# Evidence that Eddington ratio depends upon a supermassive black hole's mass and redshift: implications for radiative efficiency

Yash Aggarwal ⋆

*Lamont–Doherty Earth Observatory (Retired) Palisades, NY 10965, USA*

## ABSTRACT

Presently, it is unclear whether the Eddington ratio ($\lambda$) and radiative efficiency ($\varepsilon$) depend on a supermassive black hole's (SMBH's) redshift $z$ and mass $M_{BH}$. We attempt to resolve this issue using published data for 132 000 SMBHs with $M_{BH} \geq 10^7$ M$_{sun}$ (solar masses) at $\sim 0.1 < z < 2.4$ covering $\sim 10$ billion years of cosmic time, with $M_{BH}$ determined using Mg II lines and bolometric luminosities $L_{bol}$ based on a weighted mean of $L_{bol}$ from two or more monochromatic luminosities and a single uniformly applied correction factor. The SMBHs are sorted into seven $M_{BH}$ bins separated from each other by half an order of magnitude. The $\lambda$ and $z$ data in each bin are subjected to spline regression analysis. The results unambiguously show that for similar-size SMBHs, $\lambda$ decreases as $z$ decreases, and that for a given redshift, larger SMBHs have a lower $\lambda$. These findings require that either an SMBH's accretion rate and/or its radiative efficiency be a function of $z$ and $M_{BH}$ and, in the context of the Bondi accretion model, imply that radiative efficiency is an inverse function of $z$ and $M_{BH}$. These findings suggest that SMBHs become less efficient (higher $\varepsilon$) in accreting gases as the ambient gas density decreases with $z$ and that larger SMBHs are more efficient (lower $\varepsilon$) than smaller ones. The results leave little doubt that the current widespread practice of assigning $\varepsilon$ a standard value is untenable and gives erroneous estimates of accretion rates and growth times of SMBHs.

## 1. INTRODUCTION

The Eddington ratio ($\lambda$) of a supermassive black hole (SMBH), or the ratio of its bolometric luminosity ($L_{bol}$) to its Eddington luminosity ($L_{EDD}$) and a function of its accretion rate $\dot{M}$, radiative efficiency $\varepsilon$, and mass $M_{BH}$, is a fundamental parameter that governs its growth through cosmic times. It, however, remains uncertain whether $\lambda$ is a function of a black hole's (BH's) mass and/or redshift $z$ as exemplified by the following case studies. Shen et al. (2019) did not find a significant difference in the distribution of $\lambda$ for their high-$z$ (>5.65) sample of 50 quasars and a luminosity-matched control sample of quasars at $z = 1.5$–2.3, whereas Willott et al. (2010) found a marked difference in the distribution of $\lambda$ at $z = 2$ and $z = 6$, with the latter quasars accreting in general at higher $\lambda$ than the former. Furthermore, Shen et al. (2008) obtained mean values of $\lambda$ ranging from $\sim 0.079$ to 0.25 with typical widths of <0.3 dex for active galactic nuclei (AGNs) in different luminosity bins at $0.1 \lesssim z \lesssim 4.5$. In contrast, Kelly et al. (2010) found that compared to previous work their inferred $\lambda$ for a sample of 9886 quasars at $1 < z < 4.5$ corrected for completeness is shifted towards lower values of $\lambda$ peaking at $\sim 0.05$ with a dispersion of $\sim 0.4$ dex. In fact, Schulze & Wisotzki (2010) found a $z$-dependence of $\lambda$ in the Shen et al. (2008) data, noting that restricting their sample to $z \leq 0.3$ gives a lower mean $\lambda$ of $\sim 0.067$ with a dispersion of $\sim 0.43$ dex. Furthermore, Trakhtenbrot & Netzer (2012) found a steep rise in $\lambda$ with $z$ up to $z \simeq 1$ with smaller BHs accreting at higher values of $\lambda$. Perhaps the strongest statistical evidence thus far in favour of a probable decrease in $\lambda$ with $z$ is the observation that the $\lambda$ for high-$z$ quasars are almost all $\geq 0.1$ (e.g. Shen et al. 2019), whereas $\lambda$ for those at $1 < z < 2$ determined by Suh et al. (2015) cover a much wider range down to 0.001. Besides being important in its own right, investigating Eddington ratio's dependence on $z$ may shed light on whether radiative efficiency $\varepsilon$ is also a function of $z$. Radiative efficiency, a key input in estimating accretion rate from bolometric luminosity, is currently almost universally assumed to be 0.1, its canonical value.

Ascertaining whether $\lambda$ is a function of $z$ has been problematic for several reasons. First, multiple uncertainties affect the determination of $\lambda$ resulting from uncertainty in the monochromatic luminosity used, the bolometric correction factor applied, and the mass $M_{BH}$ of an SMBH. Secondly, it is difficult to cross-compare the results of two studies that use different bolometric correction factors. And since Eddington luminosity is a function of $M_{BH}$, $\lambda$ could also be a function of $M_{BH}$ that may obscure the dependence of $\lambda$ on $z$. The available $\lambda$ data for AGNs at $z > 3$ are not large enough to separate BHs into narrow $M_{BH}$ bins and meaningfully investigate the dependence of $\lambda$ on $z$. Kozlowski's (2017) catalogue of properties of 280 000 AGNs in the Sloan Digital Sky Survey, however, provides a uniform subset of $\sim 132\,000$ AGNs that largely overcomes the preceding handicaps. Using this subset, we investigate the dependence of $\lambda$ on $z$ and $M_{BH}$. The results are then analysed taking into account the various parameters that define $\lambda$ and infer the dependence of a BH's radiative efficiency, if any, on its mass and redshift.

⋆ E-mail: haggarwal@hotmail.com

## 2. OBSERVATIONAL DATA AND METHODS

Kozlowski's (2017) catalogue of properties of 280 000 AGNs at ∼0.1 < $z$ < 2.4 contains a subset of ∼132 000 AGNs with $M_{BH}$ > $10^7$ M$_{sun}$ (solar masses) determined using the more reliable Mg II lines and $L_{bol}$ based on a weighted mean of $L_{bol}$ from two or more monochromatic luminosities. The uncertainties in $L_{bol}$ thus determined are reduced and the reported uncertainties are generally small. The same correction factor is used in determining $L_{bol}$ and uniformly applied. Thus, the disparity arising from the use of different correction factors is eliminated. The reported dispersion in $\lambda$ values within a given $\Delta z$ is small, except for AGNs at $z < 0.5$ arising from the relative paucity of data at very low redshifts. The subset was divided into seven narrow $M_{BH}$ bins to investigate the dependence of $\lambda$ on $z$ and $M_{BH}$. The seven bins of AGNs with $M_{BH}$ in solar masses M$_{sun}$ are 674 AGNs with 1–3 × $10^7$ M$_{sun}$, 5236 AGNs with 6–10 × $10^7$ M$_{sun}$, 36 871 AGNs with 1–3 × $10^8$ M$_{sun}$, 28 304 AGNs with 6–10 × $10^8$ M$_{sun}$, 28 799 AGNs with 1–3 × $10^9$ M$_{sun}$, 2069 AGNs with 6–10 × $10^9$ M$_{sun}$, and 523 AGNs with 1–3 × $10^{10}$ M$_{sun}$. The seven bins are separated in $M_{BH}$ from adjacent bins by approximately half an order of magnitude. The $\lambda$ versus $z$ data in each bin were subjected to spline regression analysis (Belisle 1992), a method of piecewise polynomial regression in which data within subsets of $z$ are made to fit with separate models if necessary and the points at which different models are applied are called knots. This method is particularly useful where the relationship between the independent and dependent variables is not known a priori and may not be adequately captured by a single polynomial function.

## 3. RESULTS

The results are plotted in Fig. 1 that show a spline regression plot of the Eddington ratio $\lambda$ on a log scale as a function of $z$ (linear scale) using ggplot2 by Wickham (2016) covering a time span of almost 10 billion years. Each line represents the mean value of $\lambda$ as a function of $z$, and the grey shaded area shows the 95 per cent confidence interval. The probability that $\lambda$ is within the shaded area depends upon the number of data points within each subset of redshift. The number of data points or AGNs decreases at lower redshifts and hence there is in general a larger uncertainty at $z < 0.5$. We note that some of the curves in Fig. 1 have knots, justifying posteriorly the use of spline regression. On the basis of Fig. 1, we can draw the following generalized observations.

First, for a given $M_{BH}$, $\lambda$ decreases as $z$ decreases, but the rate of decrease apparently depends upon $M_{BH}$. The decrease in $\lambda$ with $z$ is unambiguous in Fig. 1 for all $M_{BH}$ bins except for the two with the most massive BHs. For example, the decrease in $\lambda$ from say $z = 2.2$ to $z = 0.75$ is ∼10 fold for the least massive 1–3 × $10^7$ M$_{sun}$ group (black line), and a factor of ∼4 for the 1–3 × $10^8$ M$_{sun}$ (red line) and the 1–3 × $10^9$ M$_{sun}$ (green line) groups. Similar decreases in $\lambda$ with $z$ can be observed for two groups represented by broken black and red lines. Furthermore, we note that the distribution of $\lambda$ values for 50 quasars at $z > 5.6$ shown in fig. 9 of Shen et al. (2019), of which nearly all except a few are >$10^9$ M$_{sun}$, define a broad peak with a median value of 0.32, whereas the BHs in the blue and green bins with $M_{BH} \geq 10^9$ M$_{sun}$ in Fig. 1 at say $z \sim 1$ have $\lambda \leq \sim 0.05$ or ∼6 times lower than the average $\lambda$ for similar-size BHs at $z > 5.6$. The results for the most massive groups in Fig. 1 (blue and broken green line) are, however, ambiguous given the much larger dispersion in $\lambda$ and the fact that $\lambda$ for these groups are largely <0.05 or close to the limiting value of $\lambda = 0$.

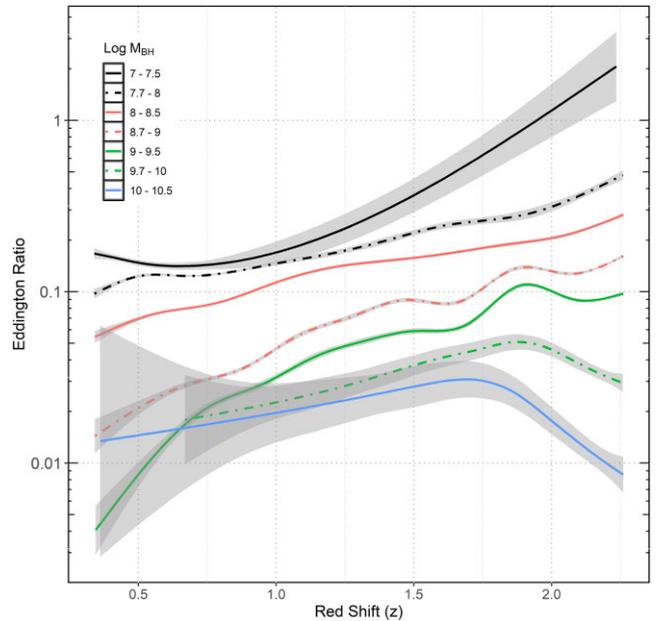

**Figure 1.** Spline regression plot of Eddington ratio $\lambda$ (log scale) versus redshift $z$ (linear scale) using ggplot2 by Wickham (2016). The data are taken from Kozlowski's (2017) catalogue separated into seven $M_{BH}$ mass bins: log $M_{BH}$ = 7–7.5, 7.7–8, 8–8.5, 8.7–9, 9–9.5, 9.7–10, and 10–10.5 separated from adjacent bins by approximately half an order of magnitude. The bins have the following number of BHs: 674, 5236, 36 871, 28 304, 28 799, 2069, and 523, respectively. Each line gives the median value of $\lambda$ as a function of $z$, and the grey area around it shows the 95 per cent confidence interval. Note that there are relatively few data points for the bin with the most massive BHs at $z < 1$. In contrast, the bin with Log $M_{BH}$ = 9-9.5 at $z \sim 0.4$ has 35 data points.

Secondly, for a given redshift, larger BHs in general have a lower $\lambda$. For example, at $z = 1.5$, $\lambda$ decreases from ∼0.4 for the least massive group to ∼0.18 for the red group, to ∼0.06 for the green group, and <0.04 for the most massive blue group (Fig. 1), a decrease in $\lambda$ by approximately an order of magnitude from the least to the most massive group.

## 4. IMPLICATIONS

Having established that $\lambda$ is a function of an SMBH's $z$ and $M_{BH}$, we now seek to assess the implications of these findings for radiative efficiency. By definition, $\lambda$ is the ratio of a BH's bolometric luminosity ($L_{bol}$) to its Eddington luminosity ($L_{EDD}$). The denominator $L_{EDD}$ is solely a function of and $\propto M_{BH}$. Hence, $\lambda$ is $\propto L_{bol}/M_{BH}$ and tracks a BH's luminosity per unit BH mass as a function of $z$ and $M_{BH}$ in Fig. 1. The numerator $L_{bol}$ is a function of a BH's accretion rate $\dot{M}$ and $\varepsilon$ the radiative efficiency. The latter is conventionally defined with respect to the mass inflow rate on to a BH, and a BH's accretion rate $\dot{M}$ is smaller by a factor of $(1 - \varepsilon)$. Hence, $L_{bol} = (\dot{M}c^2)\varepsilon/(1-\varepsilon)$, where $c$ is the velocity of light, and therefore $\lambda \propto (\dot{M}/M_{BH})\varepsilon/(1-\varepsilon)$. The order of magnitude changes in $\lambda$ with $z$ and $M_{BH}$ noted above cannot plausibly be accounted for by changes in $\varepsilon$ alone. For example, an ∼10 fold decrease in $\lambda$ from $z = 2.2$ to $z = 0.75$ observed in Fig. 1 (black curve) would require a ∼10 fold decrease in $\varepsilon/(1-\varepsilon)$ if attributed to changes in $\varepsilon$ alone. Hence, $\dot{M}/M_{BH}$ or the accretion rate must be a function of $z$ and $M_{BH}$, and in any analytical model accretion rate should explicitly or implicitly be a function of $z$ and $M_{BH}$ besides other parameters. In the Bondi accretion model, the mass-inflow rate $\dot{M}_B$, besides being a function of the sound speed, is $\propto M_{BH}^2 \rho$, where $\rho$ is the gas density far from the BH (Bondi 1952). The ambient gas density $\rho$ scales as $(1 + z)^3$ in the standard cosmological model and observational data for star-forming galaxies show that electron density decreases markedly with redshift (Rebecca et al. 2021). Hence, in the Bondi model, the mass-inflow rate per unit BH mass or $\dot{M}_B/M_{BH} \propto M_{BH}(1 + z)^3$ or a function of $z$ and $M_{BH}$ as required. The gas flow in the classical Bondi model is adiabatic, spherically symmetric, and steady state; and some workers (e.g. Gaspari et al. 2013, and references therein) have argued that it does not take into account physical processes such as radiative cooling and turbulence that could affect BH accretion rate, and therefore the Bondi assumptions should be relaxed. One such modification to the Bondi prescription has been proposed by Hobbs et al. (2012) for the case of efficient cooling and significant contribution of the surrounding halo to the total gravitational potential. The major change is that $M_{BH}$ in the Bondi prescription is replaced in the modified version by gas mass within a BH's capture radius, and the relative velocity between the BH and gas (zero in the case of the classical Bondi model) is replaced by the velocity dispersion for the external potential (see equation 8, Hobbs et al. 2012). And, in the torque-limited accretion model, the accretion rate is very weakly dependent on or almost independent of $M_{BH}$, but depends strongly on the total stellar and gas disc mass (see equation 2, Angles-Alcazar et al. 2013).

In view of the preceding observation and analyses of accretion models, the Bondi model appears to be the logical choice to investigate the dependence of radiative efficiency $\varepsilon$ on $z$ and $M_{BH}$. The Bondi prescription has been extensively used, including in those cases such as Messier 87, NGC 3115, and NGC 1600 for which the observed luminosities are orders of magnitude lower than those predicted from Bondi accretion rates and an assumed radiative efficiency (e.g. Russell et al. 2015; Wong et al. 2013; Runge and Walker 2021). Hence, we will first assess the implications for $\varepsilon$ in the context of the classical Bondi prescription and then revisit the implications with some modifications to the classical model. Under the assumptions of spherical symmetry and negligible angular momentum, the classical Bondi accretion rate $\dot{M}_B \propto M_{BH}^2 \rho/C_s^3$ (Bondi 1952), where $\rho$ and $C_s$ are, respectively, the density and sound speed at the BH's Bondi radius. And, since $C_s^2 \propto T$ the gas temperature, $\rho$ scales as $(1 + z)^3$ in the standard cosmological model and apparently does not depend upon $M_{BH}$ as discussed later, we can express $\lambda$ or the bolometric luminosity per unit $M_{BH}$ as follows:

Eddington ratio $\lambda \propto (M_{BH}/T^{3/2})(1 + z)^3 \varepsilon/(1 - \varepsilon)$. (1)

Temperature $T$ scales as the ratio of a BH's $M_{BH}$ to its Bondi radius (see e.g. Russell et al. 2015) and is a function of $M_{BH}$. Thus, for a given $M_{BH}$, $\lambda \propto (1 + z)^3 \varepsilon/(1 - \varepsilon)$ and hence $\lambda$ should decrease as $z$ decreases. This in fact is observed and amply demonstrated to be the case (Fig. 1). More importantly, however, the implication is that $\varepsilon/(1 - \varepsilon) \propto \lambda/(1 + z)^3$ for similar-size BHs irrespective of whether $\rho$ is a function of $M_{BH}$. Using this scaling relationship, we can estimate the relative change in $\varepsilon$ with $z$ for a given $M_{BH}$. For this purpose, the best publicly available data at high redshifts (say $z > 5$) are for the green group of BHs in Fig. 1 with $1$–$3 \times 10^9$ M$_{sun}$. Table 3 in Shen et al. (2019) that lists properties of 50 high-$z$ AGNs contains 17 BHs with redshifts within a narrow range of $z = 6 \pm 0.2$ and $M_{BH}$ comparable to those in the green bin in Fig. 1. These high-$z$ AGNs have a mean $\lambda$ value of $\sim 0.4$, whereas those in Fig. 1 have $\lambda \sim 0.03$ at $z = 1$. Applying the scaling relation $\varepsilon/(1 - \varepsilon) \propto \lambda/(1 + z)^3$, we get an increase in $\varepsilon/(1 - \varepsilon)$ by a factor of 3.2 from $z = 6$ to $z = 1$. And assuming $\varepsilon = 0.1$ at $z = 6$, we get radiative efficiency $\varepsilon = \sim 0.26$ at $z = 1$, or an increase in $\varepsilon$ by a factor of $\sim 2.6$ from $z = 6$ to $z = 1$ for BHs with $1$–$3 \times 10^9$ M$_{sun}$. For the same group, the increase in $\varepsilon/(1 - \varepsilon)$ from say $z = 2.25$ to $z = 1$ in Fig. 1 is by a factor of $\sim 1.28$ and the corresponding increase in $\varepsilon$ is relatively smaller or by a factor of $\sim 1.2$ from $z = 2.25$ to $z = 1$.

Publicly available $\lambda$ data for the other groups at high redshifts ($z > 5$) are not sufficient to meaningfully assess the change in $\varepsilon/(1 - \varepsilon)$ from very high to low $z$. A similar analysis, however, of the data in Fig. 1 for the other groups shows the following. Note that an increase in $\varepsilon/(1 - \varepsilon)$ implies an increase in $\varepsilon$ the radiative efficiency. For the red group in Fig. 1, the increase in $\varepsilon/(1 - \varepsilon)$ is roughly by a factor of $\sim 1.8$ from $z = 2.25$ to $z = 1$. The $\lambda$ values for the most massive BHs (blue curve) have large dispersion and are close to the limiting value of zero, and hence a change in $\varepsilon/(1 - \varepsilon)$ if any cannot possibly be discerned. Similarly, for the least massive group (black curve) the uncertainty in $\lambda$ increases as $z$ increases, making it difficult to meaningfully ascertain a change if any in $\varepsilon/(1 - \varepsilon)$ from $z = 2.25$ to $z \sim 1$. Note that $\lambda$ values are plotted on a log scale. For the second-least massive group (broken black curve), however, that is approximately half an order of magnitude more massive than the least massive (black curve), $\varepsilon/(1 - \varepsilon)$ increases by a factor of $\sim 1.4$ from $z = 2.25$ to $z = 1$. Lastly, for the broken-red line group, $\varepsilon/(1 - \varepsilon)$ increases from $z = 2.25$ to $z = 1$ by a factor of $\sim 1.28$. We stress that the preceding relative increases in $\varepsilon/(1 - \varepsilon)$ and hence in $\varepsilon$ are rough estimates, but clearly establish that in general radiative efficiency $\varepsilon$ increases as $z$ decreases and that the change in $\varepsilon$ apparently depends on a BH's mass. The finding suggests that $\varepsilon$ is probably an inverse function of the ambient gas density, which is consistent with the suggestion by Wyithe & Loeb (2012) that when a BH is embedded in dense gas the radiation pressure is less effective.

To assess the dependence of $\varepsilon$ on $M_{BH}$, we need to first ascertain the dependence of $T$ on $M_{BH}$. In doing so, it is worth exploring whether density $\rho$ at the Bondi radius depends on $M_{BH}$. There are only three SMBHs for which $T$ and $\rho$ have been determined using *Chandra* X-ray observations, and hence any generalized conclusions drawn from the following analyses of their data should be considered somewhat tentative. Note, however, that the preceding inferences and conclusions stand on their own. All three SMBHs are observed at similar redshifts at $z < 0.01$. The galaxy NGC 3115 harbours an SMBH of $2 \times 10^9$ M$_{sun}$ (Kormendy et al. 1996) for which Wong et al. (2013) obtained a well-constrained $T = 0.3$ keV. The galaxy NGC 1600 harbours an SMBH of $17 \pm 1.5 \times 10^9$ M$_{sun}$ (Thomas et al. 2016) for which Runge & Walker (2021) obtained a $T = (1.2 + 0.15/-0.13)$ keV. Their ratio $M_{BH}/T^{3/2}$ that appears in eqation (1) normalized to $10^9$ M$_{sun}$ are, respectively, $\sim 12.17$ and $12.95 \pm 1.15$ for NGC 3115 and NGC 1600 or essentially identical despite the fact that their masses differ by approximately an order of magnitude. For Messier 87, Di Matteo et al. (2003) obtained $T = 0.8 \pm 0.01$ keV that was confirmed by Russell et al (2015) who obtained $T = (0.91 + 0.08/-0.11)$ keV with slight variations depending upon the direction of measurement (Russell et al. 2018). There are several estimates for the mass of M87 of which those by Gebhardt et al. (2011), Oldham & Auger (2016), and the Event Horizon Collaboration (2019) concur with each other and give an average of $\sim(6.8 \pm 0.8) \times 10^9$ M$_{sun}$ and hence $\sim 9.5 \pm 1.2$ for its ratio $M_{BH}/T^{3/2}$ also normalized to $10^9$ M$_{sun}$. The $M_{BH}/T^{3/2}$ for the three SMBHs are within a factor of $\sim 1.2$ of their average of $\sim 11.5$ while their $M_{BH}$ range over approximately an order of magnitude. These data, albeit limited, indicate that $\lambda$ depends little on the term $M_{BH}/T^{3/2}$ in equation (1) especially in comparison to its strong dependence on density $\rho$ that scales as $(1 + z)^3$. Hence, as a good approximation, the term $M_{BH}/T^{3/2}$ in equation (1) can be assumed to be nearly a constant and equation (1) reduces to $\lambda \propto \varepsilon/(1-\varepsilon)(1+z)^3$, assuming as in equation (1) that density $\rho$ at the Bondi radius is not a function of $M_{BH}$.

The ambient gas density scales as $(1+z)^3$. All three aforementioned SMBHs have similar redshifts. Locally, however, gas density at a BH's Bondi radius $R_B$ depends on the location of its $R_B$ and the value of the power-law index in $\rho \propto R^{-n}$ describing the decrease in $\rho$ with distance $R$ from the BH. The density profiles of NGC 1600 (Runge & Walker 2021) and NGC 3115 (Wong et al. 2013) that are well defined show that $\rho$ at their Bondi radii are almost identical despite the fact that the former is almost an order of magnitude larger than the latter. In contrast, $\rho$ for M87 that has a mass in between those of NGC 3115 and NGC 1600 is apparently higher (see Russell et al. 2015, 2018). Whether these differences reflect local conditions or are an artefact of uncertainties in the density profile of M87 is not clear. We can, however, conclude that available data, albeit limited, do not show any systematic dependence of $\rho$ at the Bondi radius on $M_{BH}$.

The conclusion or the finding that $\lambda \propto \varepsilon/(1-\varepsilon)(1+z)^3$ implies that $\varepsilon/(1-\varepsilon) \propto \lambda$ for a given redshift. The data in Fig. 1 show that for a given redshift, $\lambda$ decreases as $M_{BH}$ increases. This inverse dependence of $\lambda$ on $M_{BH}$ for a given $z$ is unambiguous for all groups at $z > 1$. The implication is that $\varepsilon$ is an inverse function of $M_{BH}$ or that more massive BHs are more efficient in accreting gases.

The above analyses and conclusions are in the context of the classical Bondi accretion in which the gas flow across the Bondi radius is constant. Let us now consider the possibility that only a fraction $\beta$ (<1) reaches the BH. In this case, equation (1) can be rewritten as follows using the approximation that $M_{BH}/T^{3/2}$ is nearly a constant:

The Eddington ratio $\lambda \propto \sim \beta\varepsilon/(1-\varepsilon)(1+z)^3$. (2)

We do not know whether the factor $\beta$ in equation (2) is a function of an AGN's redshift, $M_{BH}$, and/or the host galaxy's morphology. The largely smooth curves with few knots in Fig. 1 suggest that $\beta$ is probably not some unknown function of the morphology of the host galaxy. Let us, however, assume for the sake of argument that $\beta$ is a function of $z$ and $M_{BH}$. For a given $M_{BH}$, it is unlikely that $\beta$ is an inverse function of $z$ or that $\beta$ increases as $z$ or gas density decreases. Conversely, if $\beta$ is a direct function of $z$, then the increase in $\varepsilon/(1-\varepsilon)$ and hence in $\varepsilon$ as $z$ decreases would be even greater than that predicted by equation (1). On the other hand, it is unlikely that $\beta$ is an inverse function of $M_{BH}$. And if it is a direct function of $M_{BH}$, then equation (2) predicts that for a given $z$, the dependence of $\varepsilon$ on $M_{BH}$ would be even greater than that predicted by equation (1). In short, if $\beta$ is a function of $z$ and $M_{BH}$, then it is likely that the dependence of $\varepsilon$ on $z$ and $M_{BH}$ would be even more pronounced than that predicted by equation (1).

## 5. CONCLUSIONS

In conclusion, publicly available data for Eddington ratios $\lambda$ for many tens of thousands of SMBHs clearly show that $\lambda$ is an inverse function of a BH's mass and decreases as a BH ages or its redshift decreases. Viewed in the context of the Bondi accretion model, the implication is that the radiative efficiency $\varepsilon$ is apparently an inverse function of both redshift and mass $M_{BH}$. These findings suggest that radiative efficiency increases or that BHs become less efficient in accreting gases as gas density decreases with redshift and that larger BHs are more efficient in accreting gases than smaller ones. Quantitatively, the analysis shows that for BHs the size of $10^9$ M$_{sun}$, the radiative efficiency $\varepsilon$ at $z = 1$ should be higher by roughly a factor of $\sim 2.5$ compared to that at $z = 6$. More importantly, it is clear that the current widespread use of a standard value of 0.1 for $\varepsilon$ is unwarranted and may result in highly erroneous estimates of accretion rates of SMBHs from bolometric luminosities.

## ACKNOWLEDGEMENTS

I thank Manuel Chirouze for performing the spline regression analysis used in this study and the referee for constructive suggestions that helped improve the article.

## DATA AVAILABILITY

No new data were generated in this study.

This paper has been typeset from a Microsoft Word file prepared by the author.